\documentclass[prl,preprint]{revtex4}%
\usepackage{amsfonts}
\usepackage{amsmath}
\usepackage{amssymb}
\usepackage{graphicx}%
\providecommand{\U}[1]{\protect\rule{.1in}{.1in}}

\begin{document}
\title{}

\section{Comment on "Nonexponential Decay Via Tunneling in Tight-Binding
Lattices and the Optical Zeno Effect"}

S. Longhi [1] studied the survival probability $P(t)$ of an unstable state
coupled to a tight-binding lattice finding an exact analytical solution that
describes the nonexponential decay. When the first coupling is smaller than
the others, he shows that $P(t)$ has a natural decomposition into two terms;
one is the exponential decay, consistent with the Gamow%
\'{}%
s approach, and the other is the correction to this decay. The first purpose
of this Comment is to show that the condition imposed on the system; the site
energy of the first site is set in the center of the band, limits the
generality of the results. A decomposition based on the spectral properties of
the system removes this restriction [2]. Other point of this Comment is that,
for the weak coupling limit, the author stated that a Zeno effect occurs for
$\tau<\tau^{\ast}$, where $\tau^{\ast}$ is the smallest root of the equation
$\gamma_{eff}\left(  \tau^{\ast}\right)  =\gamma_{0},$\ and for the strong
coupling limit an anti-Zeno effect occurs for $\tau\sim\tau^{\ast\ast}$, where
$\tau^{\ast\ast}$ is the first peak of $\gamma_{eff}.$ By using a spectral
formulation we show that, the anti-Zeno effect not only requires a strong
coupling limit, also is necessary to choose the resonance energy and width
such that a destructive interference can occur. We present this interference
and give an analytical expression and physically interpretation for
$\tau^{\ast}$ and $\tau^{\ast\ast}$.

According to Ref. [1], the amplitude of the survival probability,
$c_{1}\left(  t\right)  ,$ behaves as [Eq. (11) of Ref. [1]]
\begin{equation}
c_{1}\left(  t\right)  =\sqrt{Z}\exp\left(  -\gamma_{0}t/2\right)  +s\left(
t\right)
\end{equation}
where $s\left(  t\right)  $ is the correction to the exponential term,
$\sqrt{Z}$ is a constant that depends on the normalized hopping amplitude
between the first and second site, $\Delta,$\ and $\gamma_{0}=2\Delta
^{2}\left(  1-\Delta^{2}\right)  ^{-1/2}$ is the decay rate. The formulation
developed in Ref. [2] is based on the Green function formalism, and yields to
[Eq. (10) of Ref. [2]]%
\begin{equation}
c_{2}\left(  t\right)  =a\exp\left(  -\left(  \Gamma_{0}+\mathrm{i}%
\varepsilon_{r}\right)  t\right)  +R\left(  t\right)  \label{Eq_nos}%
\end{equation}
where $a$ is a constant, $R\left(  t\right)  $ the correction to the
exponential term, $\varepsilon_{r}$ the resonance energy, and $\Gamma
_{0}=\Delta^{2}/\left(  1-\Delta^{2}\right)  \left(  1-\Delta^{2}-\left(
\varepsilon_{0}/2-1\right)  ^{2}\right)  ^{1/2}$ is the decay rate, which
include his case for the site energy of the first site in the center of the
band ($\varepsilon_{0}=2$). Note that, while for this case, the survival
probability presents consecutive dips for the long time regime, the general
situation of $\varepsilon_{0}\neq2$, the long time power law is modulated by
an oscillation with small amplitude [Eq. [32] of Ref. [2]]. This would
restrict the stated conditions [1] for the anti-Zeno effect.

The anti-Zeno effect occurs when a projective measurements of the first state
is performed with a period $\tau$ that coincides with a dip in $P(t)$. In our
description [2], such dip is consequence of a destructive interference between
the contribution from the pole of the Local Density of State (this pole is
directly connected with the resonance state created by the first site), and
the contribution from the underlying band or \textit{return} amplitude. This
dip, which occurs for a time $t=t_{R}$, is seen as a sudden and brief increase
of $\gamma_{eff}\left(  t\right)  $. Therefore, acceleration of the decay may
be observed for $\tau\sim t_{R}.$ This destructive interference is clearly
seen in the weak coupling regime, where the decomposition into two terms in
$c\left(  t\right)  $ is meaningful; for a wide range of times the exponential
term dominates, while the diffusive decay of the second term dominates for
long times, and brings out the details of the spectral structure of the
system. In the cross-over of this two behaviors, both terms have similar
amplitude, and if their phase differs on $\pi$, the destructive interference
is warranted. Following Eq. [39] of Ref. [2] with $\hbar/2\Gamma_{0}$ as the
seed of the iterative equation, we obtain
\begin{equation}
t_{R}\simeq\alpha\frac{\hbar}{\Gamma_{0}}\ln\left(  \beta\frac{B}{4\Gamma_{0}%
}\right)  ,
\end{equation}
where $B$ is the bandwidth and $\alpha,\beta\gtrsim1$ are constants that
depend on the dynamics of the environment; for a semiinfinite chain
$\alpha=2.5$ and$\ \beta\sim1.6.$

For short times the decay is quadratic (for non-divergent Hamiltonian second
moment). This lasts, in the weak coupling limit, for a time proportional to
the spectral density of the final states evaluated at the decaying state
energy $\varepsilon_{0}$ [Eq. (38) of Ref. [2]],%
\begin{equation}
t_{S}\simeq\hbar\pi N_{1}(\varepsilon_{0}).
\end{equation}
Then, for time intervals $\tau<t_{S}$ \ a recursive projective measurement
produces deceleration of the decay, or Zeno effect.

\bigskip

E. Rufeil Fiori and H. M. Pastawski

Facultad de Matem\'{a}tica, Astronom\'{\i}a y F\'{\i}sica

Universidad Nacional de C\'{o}rdoba

Ciudad Universitaria, 5000

C\'{o}rdoba, Argentina

\bigskip

[1] S. Longhi, Phys. Rev. Lett. \textbf{97}, 110402 (2006).

[2] E. Rufeil-Fiori, H. M. Pastawski, Chem. Phys. Lett. \textbf{420}, 35 (2006).

\end{document}